
\magnification=1200
\baselineskip 18pt

\centerline{\bf Electron Band Structure in a Two Dimensional Periodic Magnetic
Field} \vskip .5in

\centerline{M.C.Chang and Q.Niu}
\vskip .5in

\centerline{Department of Physics, University of Texas, Austin, TX
78712} \vskip .5in

In this paper we study the energy spectrum of a
two dimensional electron gas (2DEG) in a
two dimensional periodic magnetic field. Both a square magnetic lattice and a
triangular one are considered. We consider the general case where the magnetic
field in a cell can be of any shape. A general feature of the band structure is
bandwidth oscillation as a function of the Landau index. A triangular magnetic
lattice on a 2DEG can be realized by the vortex lattice of a superconductor
film
coated on top of a heterojunction. Our calculation indicates a way of relating
the energy spectrum of the 2DEG to the vortex structure. We have also derived
conditions under which the effects of a weak magnetic modulation, periodic or
not, may be reproduced by an electric potential modulation, and vice versa.

\vskip .3in
PACS$:$ 71.25-s, 73.20Dx, 74.60.-w
\vskip .2in
To be published in Phys. Rev. B.
\vfill\eject

\noindent{\bf{I.~INTRODUCTION}}
\vskip .5cm
The behavior of a 2DEG in a
homogeneous magnetic field modulated by a periodic electric potential or
magnetic field is a subject of some recent investigations[1--7]. In
the case of a one dimensional electric modulation (1DEM), resonance between
cyclotron radius $R_c$ and the period of modulation
$a$ results in a novel magnetoresistance oscillation[1,2]. It was found that
the electrons do not move along the trench formed by the
modulation potential when $2R_c=(\kappa-1/4)a$, and move with maximum velocity
when
$2R_c=(\kappa+1/4)a$, where $\kappa$ is an integer. Because $R_c$  changes with
the magnetic field, the magnetoresistance oscillates when the  strength of the
magnetic field increases.
This result has been explained quantum mechanically by Gerhardts ${\it
et\ al}$[1], and classically (when $R_c\gg a$) by Beenakker[2]. A variant
system
that uses a  one dimensional magnetic modulation (1DMM) instead of a 1DEM shows
similar behaviors[3]. In this case the electron drift is absent when
$2R_c=(\kappa+1/4)a$, and a maximum drift velocity along the trench is
obtained when $2R_c=(\kappa-1/4)a$.

A delicate structure in the electron energy spectrum emerges when the
modulation
is two dimensional. It is well-known that the spectrum in
a magnetic field with a two dimensional electric modulation
(2DEM) exhibits fractal behavior---the so-called Hofstadter spectrum[4].
The fragmentation of bands will suppress the part of magnetoresistance
that comes from the variation of band conductivity, and only a
weaker, diffusion-related magnetoresistance oscillation can be observed[5].
Fractal spectral structure also exists for a two dimensional magnetic
modulation (2DMM), which was studied by Wu and Ulloa in a recent study. They
also studied the collective excitations of the electron gas, and found that
it does not map out the Hofstadter spectrum exactly[7].
Although it would be interesting to observe the excitation spectrum of a 2DMM
system, there is still no attempt on an experimental realization up to the
writing of this paper.

In this paper we study the
energy spectrum of the 2DEG as a function of quasi-momentum, and show
that there is
a one to one correspondence between it and the spatial distribution of the
magnetic field. Both a rectangular magnetic lattice
and a  triangular one are considered. The latter  can be generated using a
superconductor in a vortex state. Therefore we can have a periodic modulation
on
a 2DEG by coating a superconductor film on top of a semiconductor
heterojunction.
Using the connection between the energy spectrum and the shape of the
magnetic modulation, we can use the 2DEG as
a probe for getting some basic parameters of the superconductor film.

We also explore the possible connection between electric and magnetic
modulations. There is an exact mapping at the operator level between the
Hamiltonians of a 1DMM system and a 1DEM system[9]. No such  simple mapping
exists between 2D modulation systems. However, the matrix elements of the
Hamiltonian of a 2DMM are similar in structure to those of a 2DEM. Hence, it
is possible to have a formal connection between these two, albeit in
a nontrivial way. The cases with one dimensional, two
dimensional periodic modulations, and non-periodic field distributions are
discussed.

This paper is organized as follows: In Sec.2 we
study systems with weak perturbations. In Sec.3 the exact energy
spectrum is obtained by diagonalizing the Hamiltonian matrix. In
Sec.4 a way of relating the vortex structure to the behavior of a 2DEG is
discussed. In Sec.5 we discuss the correspondence between the electric
modulations and magnetic field modulations. The last section is a
conclusion.

\vskip .5cm

\noindent{\bf{II.~WEAK MODULATION}}

\vskip .3cm

\noindent{\bf{\ \ A.~Rectangular lattice}}
\vskip .3cm

Consider a 2DEG in a homogeneous magnetic field
$B_0$ perturbed by a weak rectangular magnetic lattice $\tilde{B}(x,y)$. In
order to simplify the notation, we choose magnetic length
$\lambda=(\hbar/eB_0)^{1/2}$ to be the unit of length, $B_0$ to
be the unit of magnetic field, and $\hbar \omega_c$ to be the
unit of energy, where $\omega_c=(eB_0/m)^{1/2}$ is the
cyclotron frequency. This is the most natural choice of units in the
sense that all physical quantities are around the order
of unity in the regime of interest. The Hamiltonian can be written as
$$H={1\over 2}\left(-i{\partial\over \partial x}-y+A_x(x,y)\right)^2 +
{1\over 2}\left(-i{\partial\over \partial
y}+A_y(x,y)\right)^2.\eqno(1)$$
We have choosed the vector potential
of the homogeneous magnetic field to be ${\vec A_0}(x,y)=-y{\hat
\imath}$. ${\vec A}(x,y)=A_x(x,y){\hat \imath}+A_y(x,y){\hat \jmath}$
is the vector potential of the perturbing field.

An eigenfunction of the unperturbed Hamiltonian for the n-th
level is
$$\psi_{nk_1}(x,y)={N_n\over \sqrt{L}}e^{-ik_1x}e^{-{1\over
2}(y+k_1)^2}H_n(y+k_1),\eqno(2)$$
where $N_n$ is a normalization constant, $L$ is the size of the
system, and $k_1$ is a constant. Since the perturbed system has rectangular
symmetry, we would like to construct from (2) an unperturbed eigenstate with
the
same symmetry.  Assume $a_1$ and $a_2$ to be the lattice
constants. Discrete translation operators that commute with the Hamiltonian (1)
are the magnetic translation operators
$\tilde{T}_1,\tilde{T}_2$.
$$\eqalign{\tilde{T}_1&=T_1\cr
\tilde{T}_2&=e^{ia_2x}T_2,\cr}\eqno(3)$$
where $T_1$ and $T_2$ are the usual
translation operators. An unperturbed eigenstate that satisfies
$$\eqalign{\tilde{T}_1\Psi_{nk_1k_2}&=e^{ik_1a_1}\Psi_{nk_1k_2}\cr
\tilde{T}_2\Psi_{nk_1k_2}&=e^{ik_2a_2}\Psi_{nk_1k_2}\cr}\eqno(4)$$
is found to be
$$\Psi_{nk_1k_2}=\sum^\infty_{l=-\infty}e^{-ilk_2a_2}\psi_{n,
k_1-{2\pi l\over a_1}}.\eqno(5)$$
Eq.(4) defines the quasi-momenta $k_1$ and $k_2$, which are good quantum
numbers.  We assumed there is only one flux
quantum $\Phi_0=h/e$ per unit cell in getting Eq.(5). That is, we used
$a_1a_2=2\pi$ in the new unit system. Then, a state in a Landau level is only
coupled to states in the other Landau levels with the same $(k_1,k_2)$,
relieving the necessity of degenerate perturbation.
The case
with arbitrary magnetic flux per unit cell is more complicated, and we will
comment on it later.

The first order energy perturbation can be found by calculating the diagonal
matrix elements of the Hamiltonian on the unperturbed basis:
$$\eqalign{ {\cal
E}^{(1)}_n(k_1,k_2)&={1\over 2}\sum_{l,m=-\infty}^{\infty}
\tilde{B}(l,m)\left(L^1_n(z_{lm})+L^1_{n-1}(z_{lm})\right)
e^{-z_{lm}/2}(-1)^{lm}e^{imk_1a_1}e^{-ilk_2a_2},\cr
z_{lm}&=2\pi^2\left[({l\over a_1})^2+({m\over
a_2})^2\right],\cr}\eqno(6)$$
where $\tilde{B}(l,m)$ are the Fourier components of $\tilde{B}(x,y)$,
$$\tilde{B}(x,y)=\sum_{lm}\tilde{B}(l,m)e^{2\pi ilx/a_1}e^{2\pi
imy/a_2},\eqno(7)$$
and $L_n^1$ is the associated Laguerre polynomial ($L_{-1}^1\equiv 0$).
Notice that $\tilde{B}(x,y)$ is in units of $B_0$.  Eq.(6) can be
readily reduced to the result of a system with a one dimensional sinusoidal
perturbation in Ref.~3.

Using the asymptotic form of the associated Laguerre polynomials in the large
$n$ limit, it can be shown that, in the 4-point approximation (where all the
Fourier components are zero except $\tilde{B}(\pm 1,0)=\tilde{B}(0,\pm 1)\ne
0$), the
bandwidth is zero when $2\sqrt{n\pi}=(\kappa+1/4)\pi$, and reaches maximum when
$2\sqrt{n\pi}=(\kappa-1/4)\pi$. Therefore the widths of the energy bands in
Eq.(6)  oscillate with the Landau index.  Physically, it is clearer to
write  these two conditions as $2R_c=(\kappa\pm 1/4)a$, where the radius of
cyclotron motion $R_c\simeq \sqrt{2n}$ when $n$ is large, and
$a=\sqrt{2\pi}$. These are the same as the resonant conditions for a 1DMM.
The energy bands will be further
split into p sub-bands when there are a rational number (say $p/q$) of flux
quanta in a unit cell. A proper unperturbed wave function for such a system is
$$\Psi_{nk_1k_2}=\sum_{j=1}^pd_j
\sum^\infty_{l=-\infty}e^{-i(l+j/p)k_2qa_2}\psi_{nk_1-(l+j/p){2\pi
\over a_1}}.\eqno(8)$$
The coefficients $d_j$ are to be determined from a Harper-like
equation after we turn on the perturbation[10], more details can be found in
Ref.~7.

\vskip .3cm

\noindent{{\bf{\ \ B.~Triangular lattice}}
\vskip .3cm
An easy way to realize a two dimensional periodic magnetic modulation would
be using a superconductor vortex array with the symmetry of a
triangular lattice. We choose $A_0(x,y)=-y{\hat \imath}$ to make
our Hamiltonian explicitly translational invariant in $x$. The magnetic
translation operators that commute with the Hamiltonian of this system are
$$\eqalign{\tilde{T}_1&=g_1T_1\cr
\tilde{T}_2&=e^{i\sqrt{3}ax/2}g_2T_2,\cr}\eqno(9)$$  where $g_1=\exp(-i\pi )$,
and
$g_2=\exp(-i\pi/2)$ are gauge factors for shifting the center of a
hexagonal unit in ${\cal E}_n(k_1,k_2)$ to the origin. This will
become clearer in a later discussion.  Instead of putting $g_1$ and $g_2$ in
the
magnetic translation operators, we can put them on the right hand side of
Eq.(4). This amounts to a redefinition of the quasi-momenta.

An unperturbed eigenstate with triangular
symmetry is given by
$$\Psi_{nk_1k_2}=\sum^\infty_{r=-\infty}(-1)^{r(r-1)/
2}e^{ri(k_1/2-k_2)a}\psi_{n,k_1-{2\pi r\over a}},\eqno(10)$$ again
we assumed one flux quantum per unit cell ($\sqrt{3}a^2/2=2\pi$). The
case of one-half flux quantum per unit cell will be discussed at the
end of this section.

The first order energy perturbation is
$$\eqalign{ &{\cal E}_n^{(1)}(k_1,k_2)\cr &={1\over
2}\sum_{r,s=-\infty}^{\infty}
\tilde{B}(r,s)\left(L^1_n(w_{rs})+L^1_{n-1}(w_{rs})\right)
e^{-w_{rs}/2}(-1)^{rs}e^{is(k_1a+\pi)}e^{-ir(k_2a+{\pi})},\cr
&w_{rs}={2\pi\over \sqrt{3}}(s^2-sr+r^2)\cr}\eqno(11)$$
$\tilde{B}(r,s)$
are the Fourier components of $\tilde{B}(x,y)$,
$$\tilde{B}(x,y)=\sum_{r,s}
\tilde{B}(r,s)e^{  i(x{\hat{\bf\imath}}+y{\hat{\bf\jmath}})
\cdot (r{\hat{ b_1}}+s{\hat{ b_2}})  },\eqno(12)$$
where $\hat{b_1}=(2\pi/a)({\hat \imath}-{\hat \jmath}/\sqrt{3})$
and $\hat{b_2}=(2\pi/a)(2{\hat \jmath}/\sqrt{3})$
are unit vectors of the reciprocal lattice.

If we let all of the Fourier components be zero except six of them
(we will call this a 6-point approximation):
$\tilde{B}(\pm1,0)$, $\tilde{B}(0,\pm1)$, and $\tilde{B}(\pm1,\pm1)$, which are
all
equivalent to each other. Then we have
$$\tilde{B}(x,y)=2\tilde{B}(1,0)\left[2\cos({2\pi\over a}x)\cos({2\pi\over
\sqrt{3}a}y)+\cos({2\pi\over \sqrt{3}a}2y)\right]\eqno(13)$$
The
corresponding energy perturbation is
$${\cal
E}_n(k_1,k_2)=-e^{-\pi/\sqrt{3}}\tilde{B}(1,0)(L^1_n+L^1_{n-1})
{\left[\cos(ak_1)+\cos(ak_2)+\cos(a(k_1-k_2))\right]}.\eqno(14)$$
Notice that
$k_1$ and $k_2$ are along non-orthogonal directions $\tilde{T}_1$ and
$\tilde{T}_2$, we need to make a transformation
$$\eqalign{k_1&=k_x\cr k_2&={k_x\over 2}+{\sqrt{3}\over
2}k_y\cr}\eqno(15)$$
to put ${\cal E}_n$ on an orthogonal
basis $k_x$ and $k_y$. See Fig.1 for plots of $\tilde{B}(x,y)$ and
${\cal E}_0(k_x,k_y)$, where $\tilde{B}(1,0)=0.05$. Notice that the magnitude
of
variation of $\tilde{B}(x,y)$ is $9\tilde{B}(1,0)$ (from $-3\tilde{B}(1,0)$ to
$6\tilde{B}(1,0)$),
while that of ${\cal E}_0(k_x,k_y)$ is
$4.5e^{-\pi/\sqrt{3}}\tilde{B}(1,0)$, which is a much smaller
variation. It appears that the electrons sense the average magnetic field
more than they sense the fluctuation. The plot in ${\cal E}_0(k_x,k_y)$ will
be shifted to the right by $\pi$ if both factors $g_1$ and $g_2$ in Eq.(9)
are identity.

In a real physical situation, the triangular array is formed by a
superconductor vortex state, where the magnetic flux per plaquette is
$\Phi_0/2$ instead of $\Phi_0$. In this case the magnetic translation
operators defined in Eq.(3) do not commute ($T_1T_2=-T_2T_1$), and
the quasi-momenta in Eq.(4) are no longer good quantum numbers.
We have to choose a unit cell that consists of two magnetic plaquettes to have
a mutually commuting set of $H,T_1$ and $T_2$. Assume $a$ is the lattice
constant
of the magnetic plaquette, then  there is one flux quantum $\Phi_0$ within the
area $(\sqrt{3}/2)a_1a_2$ of the unit cell, where $a_1=2a$, $a_2=a$. The
energy
perturbation for this system is similar to Eq.(11), but with two differences:
Firstly, $(-1)^{rs}$ is replaced by $(-1)^{2rs}$, and thus can be dropped.
Secondly, $k_1a$ in the exponent is replaced by $k_1a_1$, and $k_2a$ by
$2k_2a_2$. Hence there are two identical regions in the first Brillouin
zone. This
is related to the following fact: Instead of choosing $a_1=2a$, we can
choose $a_2=2a$, and $a_1=a$. The overlapping region of these two different
Brillouin zones have the same energy spectrum because we are free to
enlarge our unit cell in either ways. The dispersion surfaces in the other
half regions of either Brillouin zones just replicate the energy spectrum
in the overlapping region. The existence of two identical regions in a
Brillouin zone is related to the two-fold degeneracy of such a system.

\vskip .5cm

\noindent{\bf{III.~EXACT ENERGY SPECTRUM}}
\vskip .3cm
The exact energy spectrum for an electron in a magnetic field with strong
modulation can be found by diagonalizing the Hamiltonian (1) numerically.
For a square magnetic lattice, the result of diagonalization in the 4-point
approximation is shown in Fig.2. The
energies from the perturbation calculation are lower than the exact
energies since a positive--definite
contribution from the quadratic terms of the vector potential is
neglected in calculating Eq.(6). Notice that the magnetic modulation
is of the same strength as the homogeneous component, but the energy shift is
just a fraction of  $\hbar\omega_c$.

Matrix elements of the Hamiltonian (1) with triangular
modulation are listed below for
reference. Write the Hamiltonian as $H=H_0+H_1+H_2,$
where $H_1$ is linear in $A$, and $H_2$ is quadratic in $A$. The matrix
elements of $H_1$ on the unperturbed basis (10) are found to be
$$\eqalign{ &{\langle n'|H_1(k_1,k_2)|n\rangle }_{(n'\ge
n)}\cr
&={1\over 2}\sum_{r,s} \tilde{B}(r,s)\left(L_n^{n'-n}(w_{rs})
-\left[{n'+n\over w_{rs}}L_n^{n'-n}(w_{rs})
-{2n'\over
w_{rs}}L_{n-1}^{n'-n}(w_{rs})\right]
\right) G_{n'n}(r,s).\cr}\eqno(16)$$
The matrix elements of $H_2$ are
$$\eqalign{&{\langle
n'|H_2(k_1,k_2)|n\rangle }_{(n'\ge n)}\cr  &=\sum_{{r,s\atop
r',s'}}{\left[
A_x(r,s)A_x(r',s')+A_y(r,s)A_y(r',s')\right] }L_n^{n'-n}
(w_{r+r',s+s'})G_{n'n}(r+r',s+s'),\cr}\eqno(17)$$
where the function $G_{n'n}$ is defined to be
$$G_{n'n}(r,s)=\sqrt{n!\over
n'!}(\sqrt{2}\pi)^{n'-n}{\left({ir\over a}+{2s-r\over
\sqrt{3}a}\right)}^{n'-n}e^{-w_{rs}/2}(-1)^{rs}e^{is(k_1a+\pi)}
e^{-ir(k_2a+\pi)}.\eqno(18)$$
The vector potential ${\vec A}(r,s)$ is related to ${\tilde {\vec B}}(r,s)$ by
${\tilde{\vec B}}(r,s)=i{\vec q}\times{\vec A}(r,s),$
where ${\vec q}=r{\hat{ b_1}}+s{\hat{ b_2}}\ne 0$.
Using the Coulomb gauge condition, we have
$${\vec A}(r,s)=i{{\vec q}\over |{\vec
q}|^2}\times {\tilde{\vec B}}(r,s).\eqno(19)$$
Under the 6-point approximation, there is only one independent
component, say $A_x(1,0)$, of the vector potential, which is related to
${\tilde
B}(1,0)$ by $A_x(1,0)=(\sqrt{3}a/8\pi i){\tilde B}(1,0)$.
The energy spectrum for a system with
a 6-point approximation is shown in Fig.3, the modulation
strength $\tilde{B}(1,0)$ is chosen such that the minimum of the total magnetic
field $B(x,y)=B_0+\tilde{B}(x,y)$ is zero.
Numerical calculation shows that the transition amplitudes between different
Landau levels  oscillate with respect to the level separation $\Delta n$ when
$\Delta  n$ is small, and drop quickly
to zero beyond $\Delta n\simeq 10$. The asymptotic behavior goes
roughly like
$1/(\Delta n)!$. We were careful to choose the rank of the matrix being
diagonalized  large enough to ensure the convergence of our result.

When the variation of magnetic field is large, adjacent energy
levels begin to mix with each other. This can be seen from Fig.4. The
modulation
part is nine times larger than the homogeneous part, therefore in some region
of
the unit cell the total magnetic field reverses the direction.  Notice that
these levels in general do not touch. This conforms to a theorem by Von
Neumann
and Wigner, which states that, it needs to vary at least three parameters for a
generic Hamiltonian to have an accidental degeneracy[11].

Aharonov and Casher once showed that the ground state energy of a
spin-1/2 particle in a nonuniform magnetic field is zero[12]. This result
is checked numerically in this periodic case by adding a spin
interaction term $(\sigma_z/2)B(x,y)$ to the  Hamiltonian (1). Since (10) is
not a
very good basis for this new  Hamiltonian, it is necessary to
diagonalize a larger Hamiltonian matrix to get a
correct result. The lowest dispersion curve we get for this
Hamiltonian with spin term is found to be flat and zero with negligible
deviation, which agrees very well with Aharonov and Casher's result.

\vskip .5cm

\noindent{\bf{IV.~ENERGY SPECTRUM AND VORTEX STRUCTURE}}
\vskip .3cm

It is not hard to see that, when the energy perturbation is small, we can
determine $B(x,y)$ with the help of Eq.(11) and Eq.(12) {\sl if the energy
spectrum ${\cal E}_n(k_1,k_2)$ over the entire  first Brillouin zone can be
mapped out precisely}. Therefore,  in principle, the 2DEG beneath the
superconductor film can be a probe for measuring the vortex structure.
However, this is a difficult experimental task compared to other methods like
neutron diffraction, scanning tunneling microscopy[13] and electron wave
holography[14]. Angle Resolved Photoemission Spectroscopy (ARPES) can be a tool
of measuring the band structure, but it can barely probe beyond $10 \AA$ below
the surface, while
our 2DEG is buried down in the heterojunction. Besides, the electron density of
a conventional 2DEG ($\simeq 10^{11} \rm{\ cm^{-2}}$) is too low for ARPES to
have the precision we need. In spite of these difficulties, at least the
bandwidth can be measured with confidence, and this can give us some
information about the vortex state. In this section we will take a simple model
of vortex structure as our starting point, and demonstrate how the bandwidth of
the 2DEG may help us determine the coherence length of the superconductor.

A convenient choice for the structure of a vortex was proposed by Clem[15]. The
magnetic field distribution of a single vortex in the Clem model is
$$b(\rho)={\Phi_0\over
4\pi\lambda\xi_v}{K_0\left((\rho^2+\xi_v^2)^{1/2}/\lambda\right)
\over K_1(\xi_v/\lambda)},\eqno(20)$$
where $K_0$ and $K_1$ are the modified Bessel functions, $\rho$ is a radial
coordinate, $\lambda$ is the penetration depth, and $\xi_v$ is a variational
parameter that is of the same order as the coherence length.
$\xi_v$ and $\lambda$ are the only two parameters in this model.
The magnetic field of a vortex array is
$B_s({\vec r})=\sum_{\vec R} b({\vec r}-{\vec R}),$
where the summation is over all sites of the vortices. Therefore, the
Fourier components,
$$ \eqalign{{B_s}(r,s)&={1\over L^2}\int B_s({\vec
r})e^{i{\vec q}\cdot{\vec r}} d{\vec r}\cr
&={1\over A_{cell}}{\Phi_0\over 2\lambda}{1\over
Q}{K_1(Q\xi_v)\over K_1(\xi_v/\lambda)},\cr}\eqno(21)$$
where $A_{cell}$ is the area of a unit cell,
${\vec q}=r\hat{b_1}+s{\hat b_2}$, and $Q\equiv
(|{\vec q}|^2+\lambda^{-2})^{1/2}$.

The magnitude of oscillation of the magnetic field ${B_s}(x,y)$ at the
surface of the superconductor is attenuated through the
distance $d$ between the superconductor film and the 2DEG. Fourier
component of wave vector ${\vec q}$ is damped out by a factor
$\exp(-|{\vec q}|d)$[1]. Therefore, the Fourier components of the magnetic
field that appear in Eq.(11)
are actually $\tilde{B}(r,s)$ = ${B_s}(r,s)$ $e^{-|{\vec q}|d}$, where
$|{\vec q}|=4\pi/(\sqrt{3}a)\sqrt{r^2-rs+s^2}$. For example, if
$a=1500\ \AA$ (for $B_0=0.1$ Tesla), and $d=400\ \AA$, then
$\tilde{B}(1,0)=0.16B_s(1,0)$, $\tilde{B}(1,-1)=0.04B_s(1,-1)$. The modulation
on the 2DEG is
small unless $B_0$ is much less than 0.1 T, or $d$ is much less than 400 $\AA$.
Therefore in most of the cases the perturbation formula works well.

Assuming that the vortex array is
dense enough so that the 6-point approximation is valid, then the width of the
n-th band is
$$\Delta {\cal
E}_n=4.5e^{-\pi/\sqrt{3}}\tilde{B}(1,0)(L_n^1(2\pi/\sqrt{3})+L_{n-1}^1
(2\pi/\sqrt{3})).\eqno(22)$$
One way of measuring the bandwidth is by measuring the differential Hall
conductivity of the 2DEG, which in the collisionless limit is
$$\sigma'_{H}(E)={e\over B_0}D(E)f(E),\eqno(23)$$
where $D(E)=\sum_n\int{d^2{\vec k}\over 4\pi^2}\delta(E-{\cal E}_n({\vec
k}))$ is the density of states, and $f(E)$ is the Fermi distribution. Notice
that we have put back the real units in the above formula. This expression can
be
derived from the Kubo formula, under the assumption that the modulation is
small
compared to the average field. The effect of modulation appears through the
${\cal E}({\vec k})'s$ in the density of states and in the Fermi distribution.
Eq.(23) is the same as the classical formula $\sigma_H=\rho_e e/B_0$,
where $\rho_e$ is the density of mobile electrons.

In the presence of disorder, $D(E)$ has to be replaced by
the density of extended states $D_e(E)$, or more precisely, the density of
Chern
numbers[16].  For a homogeneous magnetic field, it has been shown that $D_e(E)$
is a delta function for each Landau levels. When the magnetic field is
modulated, we expect that the width of $D_e(E)$ be approximately equal to the
bandwidth in the weak disorder limit.  A tricky point in measuring the
bandwidth
is that we need a large $B$ to have a good resolution in a quantum Hall system,
but that tends to pack the vortex array tightly and shrink the bandwidths.
Therefore it is necessary to find a compromise between these two constraints.
A
theoretical calculation of the Hall conductivity according to (23) at
zero temperature is shown in Fig.5. It varies continuously when the Fermi
energy
is inside the band, and takes quantized values when the Fermi energy is in the
gaps. The slopes  are discontinuous at the band edges since the density of
states are discontinuous for a 2D system. How a finite disorder will change
the
extended density of states remains  an open question.

{}From the bandwidths we can get one Fourier
component ${\tilde B}(1,0)$ and hence an algebraic relation between $\xi_v$ and
$\lambda$ through Eq.(21) and Eq.(22). This is useful because the coherence
length of a type-II superconductor with large $\kappa$  is usually more
difficult to be measured than the penetration depth. Therefore, if $\lambda$ is
known beforehand, measurement of the bandwidth can give us the coherence
length.
The measured values of $\tilde{B}(1,0)$ from different bands can be used as a
consistency check of this approach. In principle it is also possible to
determine the temperature dependence of $\xi_v$ since the bandwidth will be
different when $\xi_v$ is changed by temperature.

The Hall conductivity gives more information about the band structure than just
the bandwidths. One may compare the density of states measured according to
Eq.(23) and that calculated from the theoretical model. If
the sample is clean enough, such a comparison should yield information about
the
higher Fourier components of the magnetic field, and therefore more information
about the magnetic flux lattice.

When the strength of the modulation is large, neighboring bands are mixed
due to strong oscillations (see Fig.4), also Eq.(23) for the Hall conductivity
is no longer valid.  However,  each filled band still contributes an integer
number of  $e^2/h$ to the Hall conductivity as long as it is not in touch with
the other bands (usually adjacent bands are separated by avoided crossings).
These avoided crossings (or, local gaps) can be closed by tuning $\tilde{B}$ to
a particular value $\tilde{B}^*$. It does not violate the Von Neumann--Wigner
theorem since three parameters $(k_x,k_y,{\tilde B})$ are used to induce an
accidental degeneracy. When a local gap collapses, the Hall plateau that
corresponds to that gap disappears. A new plateau emerges when the gap is
re-opened by increasing $\tilde{B}$ away from $\tilde{B}^*$. This plateau may
or
may not stick to its original value. In case that it does not, $\sigma_H$ will
jump up or down by an integer multiple of $e^2/h$[17]. This dramatic effect can
also be achieved by using an electric modulation to control the band crossings.

\vskip .5cm
\noindent{\bf{V.~CORRESPONDENCE BETWEEN GENERAL MAGNETIC AND ELECTRIC
MODULATIONS}}
\vskip .3cm
{\noindent{\bf{\ \ A.~One dimensional modulation}}
\vskip .3cm
It is well--known that the equation of motion of a 2D
electron in a constant magnetic field can be reduced to that of a 1D
electron in a parabolic electric potential.  This analogy
can be generalized when the magnetic field is modulated along one
direction[9],  where the corresponding electric potential $V(x)$  is determined
by
$$B(x)={d\sqrt{2V(x)}\over dx}.\eqno(24)$$
For a weak 1DMM with $B(x)=1+\tilde{B}\cos (2\pi/a x)$, the
corresponding electric potential
$$V(x)\simeq {(x-c)^2\over 2}+\tilde{B}(x-c){a\over
2\pi}\sin {2\pi\over a} x.\eqno(25)$$
Besides the "all-magnetic" and "all-electric" systems, it is also possible
to transform the term $(x-c)^2/2$ in Eq.(25) back to a homogeneous magnetic
field, and  keep the second term intact. This is equivalent to
a mixed system with a magnetic field plus a modulating electric field.
Following Beenakker's analysis on the drift of the guiding center of an
electron
in a 1DEM[2], the mean square drift velocity in the classical limit $R_c\gg a$
is
found to be
$$\langle
v_d^2\rangle=(\tilde{B}{a\over 2\pi})^2{R_c\over a}\cos^2({2\pi R_c\over
a}-{3\pi\over 4}),\eqno(26)$$
which is the same as the one in Xue's paper[3].
It has maxima under the resonant condition $2R_c=(\kappa-1/4)a$,

\vskip .3cm
\noindent{\bf{\ \ B.~Two dimensional periodic modulation}}
\vskip .3cm

There is no simple mapping as in Eq.(24) when the modulation is
bidirectional. However, a connection still exists when the perturbation is
small. Consider a 2DEG in a rectangular 2DEM. The
first order energy perturbation is similar to that in Eq.(6). The only
difference is that
$\tilde{B}(l,m)\left(L_n^1(z_{lm})+L_{n-1}^1(z_{lm})\right)$ is
replaced by $2V_1(l,m)L_n(z_{lm})$. Therefore, as far as the energy spectrum
goes, a magnetic modulation can be simulated using an electric potential whose
Fourier components satisfy
$$2V_1(l,m)L_n(z_{lm})
=\tilde{B}(l,m)\left[L_n^1(z_{lm})+L_{n-1}^1(z_{lm})\right]
\eqno(27)$$
It can be readily reduced to a one dimensional relation if
all the components with $l\ne 0$ are zero. Unlike the effective potential
in Eq.(25), this one is valid only in the one-band
approximation, and is different for different bands.

A curious feature of this formula is
that if a $z_{lm}$ happens to coincide with a zero
of the Laguerre polynomial, the $(l,m)$th Fourier component of the
effective potential does not exist, simply because there is no way
a $V_1(l,m)$ can contribute to the broadening of the n-th level, no
matter how large it is. Therefore, for a given Landau level with $n\ne 0$, an
electric modulation can only mimic a magnetic modulation whose Fourier
components $\tilde{B}(l,m)$ vanish when $z_{lm}$ are at zeros of $L_n$.
Conversely, a magnetic modulation can only mimic an electric modulation
whose Fourier components $V_1(l,m)$ vanish when $z_{lm}$ are at zeros of
$L_n^1+L_{n-1}^1$.

\vskip .3cm
\noindent{\bf{\ \ C.~General two dimensional field distribution}}
\vskip .3cm
It is possible to extend the above connection to a random
magnetic field. The only restriction is that the
random component be weak compared to the homogeneous component. Consider
the matrix elements of the Hamiltonian in the unperturbed basis (2) (no
requirement on one flux quantum per unit cell).
Express the magnetic field in Fourier integral instead of Fourier series since
there is no discrete translational symmetry here, then we have
$$\langle n,q-q'|H_1^B(x,y)|n,q\rangle
={1\over 2}\int dk_2\tilde{B}(q',k_2)(L_n^1+L_{n-1}^1)e^{-{1\over
4}(q'^2+k_2^2)}e^{{i\over 2}q'k_2}e^{ik_2q},\eqno(28)$$
where the argument in the associated
Laguerre polynomial is $(q'^2+k_2^2)/2$. For a general electric
potential $V(x,y)$, the matrix elements are
$$\langle
n,q-q'|H_1^E(x,y)|n,q\rangle =\int dk_2V(q',k_2)L_n e^{-{1\over
4}(q'^2+k_2^2)}e^{{i\over 2}q'k_2}e^{ik_2q}\eqno(30)$$
Since the basis {$|n,q \rangle$} form
a complete set for the n-th level, it follows that the two operators
$H_1^B$ and $H_1^E$ are equivalent in
the one-band approximation if (28) and (29) are equivalent for any $q$
and $q'$. Assume the equivalence of these two matrices, we can take inverse
Fourier transforms of integrals with respect to $q$ to get an
identity between the integrands
$\tilde{B}(L_n^1+L_{n-1}^1)\exp(-(q'^2+k_2^2)/4)\exp(iq'k_2/2)$ and
$VL_n\exp(-(q'^2+k_2^2)/4)\exp(iq'k_2/2)$. After dividing out the common
exponential part, it follows that $H_1^B=H_1^E$ if and only if
$$2L_n
V(k_1,k_2)=(L_n^1+L_{n-1}^1)\tilde{B}(k_1,k_2),\ \ \ \ \ \ \forall \
k_1,k_2,\eqno(30),$$
which is a continuous version of Eq.(27).
The discussion in the previous subsection about the
singularities in the  Fourier components applies to this case as well.
This relation might be helpful to the researchers working
on problems of random magnetic field.

\vskip .5cm
\noindent{\bf{VI.~CONCLUSION}}
\vskip .3cm
We have calculated the energy spectra of 2D electrons
in two dimensional
periodic magnetic fields. The spectra are expressed in
quasi-momenta that are good quantum numbers of the system. It is found
that, the Landau levels are broadened in an oscillatory manner with respect
to the band index. When the modulation of the magnetic field is strong,
dispersion curves are mixed but show avoided crossings. The magnetic lattice
with
triangular symmetry can be realized by using a superconductor film in a
vortex state.  A connection between the electric properties of the 2DEG and the
magnetic properties of the vortex arrays is studied. It shows a possible way
of getting information about the vortices through measurements of the
bandwidth of the 2DEG. Finally, the
connection between electric field modulation and magnetic field
modulation is discussed. We find the effective electric potentials for
weak 2D magnetic modulations. The effective potentials
differ from band to band, and do not always exist.

\vskip .5cm
{\bf ACKNOWLEDGMENTS}
\vskip .3cm
We would like to thank D.~Gavenda, J.~Markert, C.K.~Shih and T~.Wang for
helpful
discussions. This work is supported by NIST and the Welch Foundation.

\vskip .5cm
{\bf REFERENCES}
\vskip .3cm
\item {1.} R.R.Gerhardts, D.Weiss, and K.v. Klitzing, {\it Phys
Rev. Lett.} {\bf 62}, 1173 (1989);\hfil\break
R.W.Winkler, J.P.Kotthaus and k.Ploog, {\it ibid} {\bf 62}, 1177 (1989).
\item {2.} P.Vasilopoulos and F.M.Peeters, {\it Phys. Rev. Lett.}
{\bf 63}, 2120 (1989); C.Zhang and R.R.Gerhardts, {\it Phys. Rev.}
{\bf B41}, 12850 (1990); R.R.Gerhardts, {\it Phys. Rev.}{\bf B45}, 3449 (1991);
C.W.J.Beenakker, {\it Phys. Rev. Lett.} {\bf 62}, 2020 (1989).
\item {3.} D.P.Xue and G.Xiao, {\it Phys. Rev.} {\bf B45}, 5986
(1992); X.Wu and S.E.Ulloa, {\it Phys. Rev.} {\bf B47}, 7182
(1993).
\item {4.} D.R.Hofstadter, {\it Phys. Rev.} {\bf B14}, 2239 (1976).
\item {5.} H.Fang and P.J.Stiles, {\it Phys. Rev.} {\bf B41},
10171 (1990);
R.R.Gerhardts, D.Weiss and U.Wulf, {\it Phys. Rev.} {\bf B43},
5192 (1991);
Y.Tan and D.J.Thouless, {\it Phys. Rev.} {\bf B49}, 1827 (1994).
\item {6.} B.A.Dubrovin and S.P.Novikov, {\it Sov. Phys. JEPT} {\bf 52}, 511
(1880); {\it Sov. Math. Dokl} {\bf 22}, 240 (1980); S.P.Novikov,
{\it Sov. Math. Dokl} {\bf 23}, 298 (1981)
\item {7.} X.Wu and S.E.Ulloa, {\it Phys. Rev.} {\bf B47}, 10028 (1993)
\item {8.} Experiments using similar set-ups have been done in the studies of
weak
localization of electrons in an inhomogeneous magnetic field[8].
See S.J.Bending, K.v. Klitzing, and K.Ploog, {\it Phys. Rev. Lett.} {\bf 65},
1060 (1990);\hfil\break  A.K.Geim, S.J.Bending, and I.V.Grigorieva, {\it Phys.
Rev. Lett.} {\bf 69}, 2252 (1992). \item {9.} F.M.Peeters and A.Matulis, {\it
Phys. Rev.} {\bf B48}, 15166 (1993).
\item {10.} D.J.Thouless, M.Kohmoto, M.P.Nightingale, and M.de
Nijs, {\it Phys. Rev. Lett.} {\bf 49}, 405 (1982).
\item {11.} J.Von Neumann and E.P.Wigner, {\it Phys. Z.} {\bf 30},
467 (1929).
\item {12.} Y.Aharonov and A.Casher, {\it Phys. Rev.} {\bf A19},
2461 (1979). The degeneracy of ground state energy for two
dimensional electrons (with spin) in an inhomogeneous magnetic
field has been generalized to the situation that the electrons are
embedded in an arbitrary 2 dimensional curved surface, with the
magnetic field being everywhere orthogonal to the surface. See
R.Alicki, J.R.Klauder, and J.Lewandowski,
preprint.
\item {13.} H.F.Hess, R.B.Robinson, and J.V.Waszczak, {\it Phys.
Rev. Lett.} {\bf 64}, 2711 (1990); F.Gygi and M.Schluter, {\it Phys.
Rev.}{\bf B43}, 7609 (1991).
\item {14.} S.Hasegawa, T.Matsuda, J.Endo, N.Osakabe, M.Igarashi,
T.Kobayashi, M.Naito,\hfil\break
A.Tonomura, and R.Aoki, {\it Phys.
Rev.} {\bf B43}, 7631 (1991).
\item {15.} J.R.Clem, {\it J. Low. Temp. Phys.} {\bf 18}, 427 (1975).
\item {16.} Y.Huo and R.N.Bhatt, {\it Phys. Rev. Lett.} {\bf 68}, 1375 (1992)
\item {17.} A detailed study has been done by M.Nielson and P.Hedegard,
preprint.

\vfil\eject

{\bf Figure captions}
\vskip .3cm

Fig.~1. (a), A plot of the magnetic field $B(x,y)$
with triangular lattice symmerty in the 6--point approximation.
(units of $B_0$)
(b), The ground state
energy of an electron in a homogeneous magnetic field modulated by the
magnetic field in (a). (units of $\hbar \omega_c$)

Fig.~2. Broadened Landau levels for a 2DEG in a homogeneous magnetic
field modulated by a square magnetic lattice. The perturbing field is
$\tilde{B}(x,y)=0.5(\cos x+\cos y)$. Dashed lines are for the first order
energy perturbation, and solid lines are for the exact energy spectrum.
Only the lowest fifteen levels are shown here. (units of $\hbar \omega_c$)

Fig.~3. Exact energy spectrum for a 2DEG in a homogeneous magnetic
field modulated by a triangular magnetic lattice in the 6--point
approximation, $\tilde{B}(1,0)=1/3$.

Fig.~4. Exact energy spectrum for a 2DEG in a homogeneous magnetic
field modulated by a triangular magnetic lattice in the 6--point
approximation, $\tilde{B}(1,0)=1$. Many of the curves at the upper left and
right are actually separated with higher resolution.

Fig.~5. Hall conductivity at zero temperature. Due to the broadened Landau
levels, the transition from one plateau to another is gradual. $\tilde{B}(1,0)$
here
is 0.2.

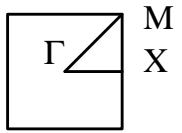
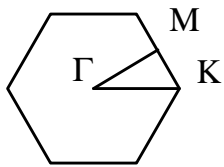
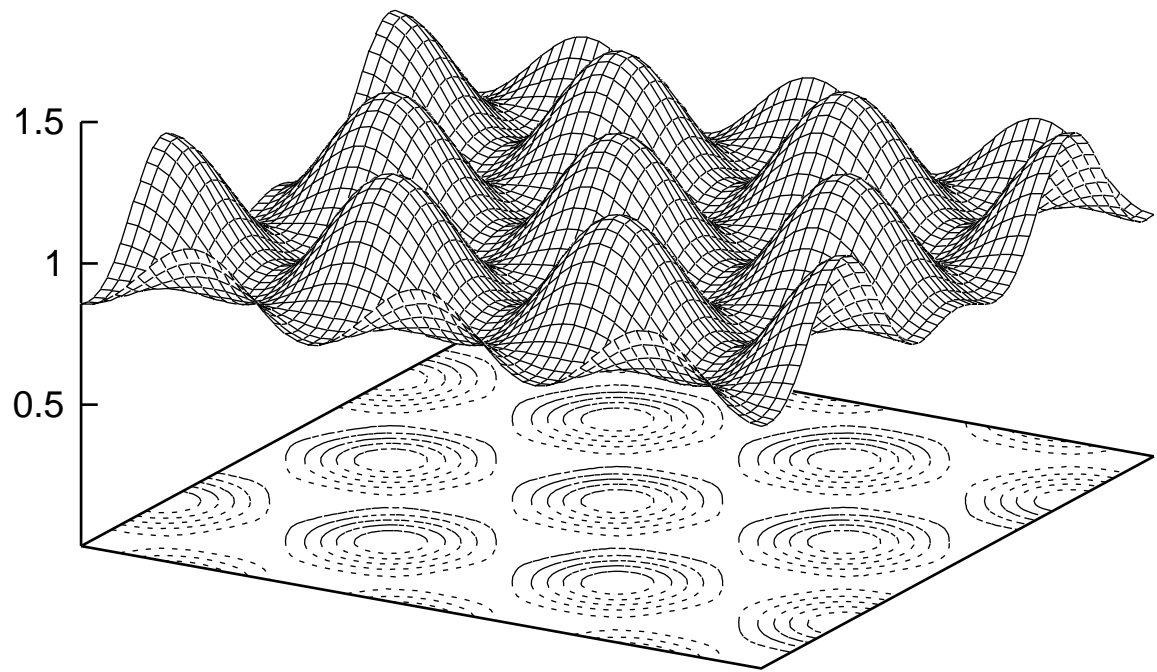
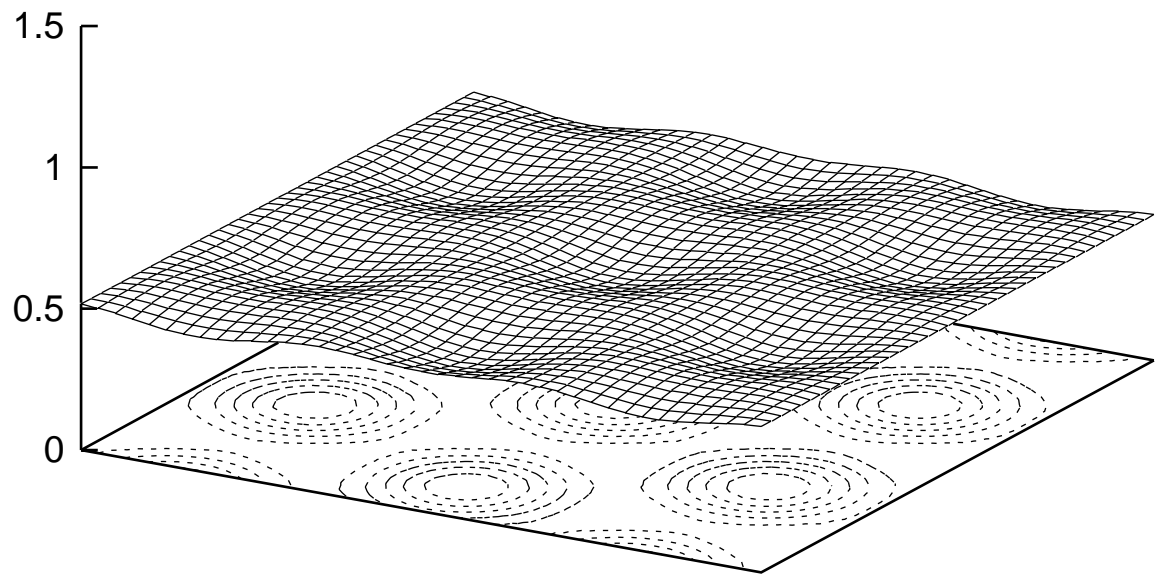
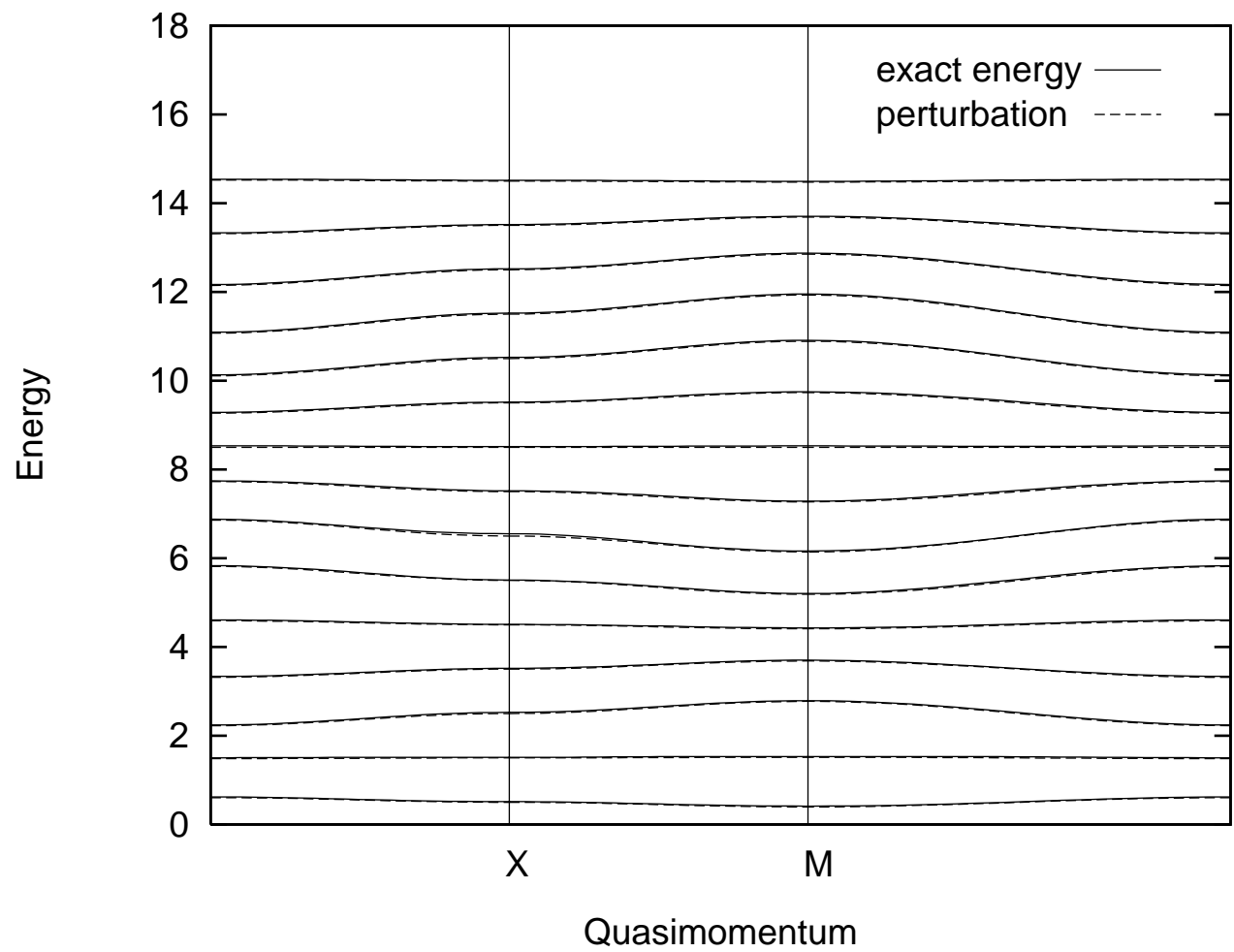
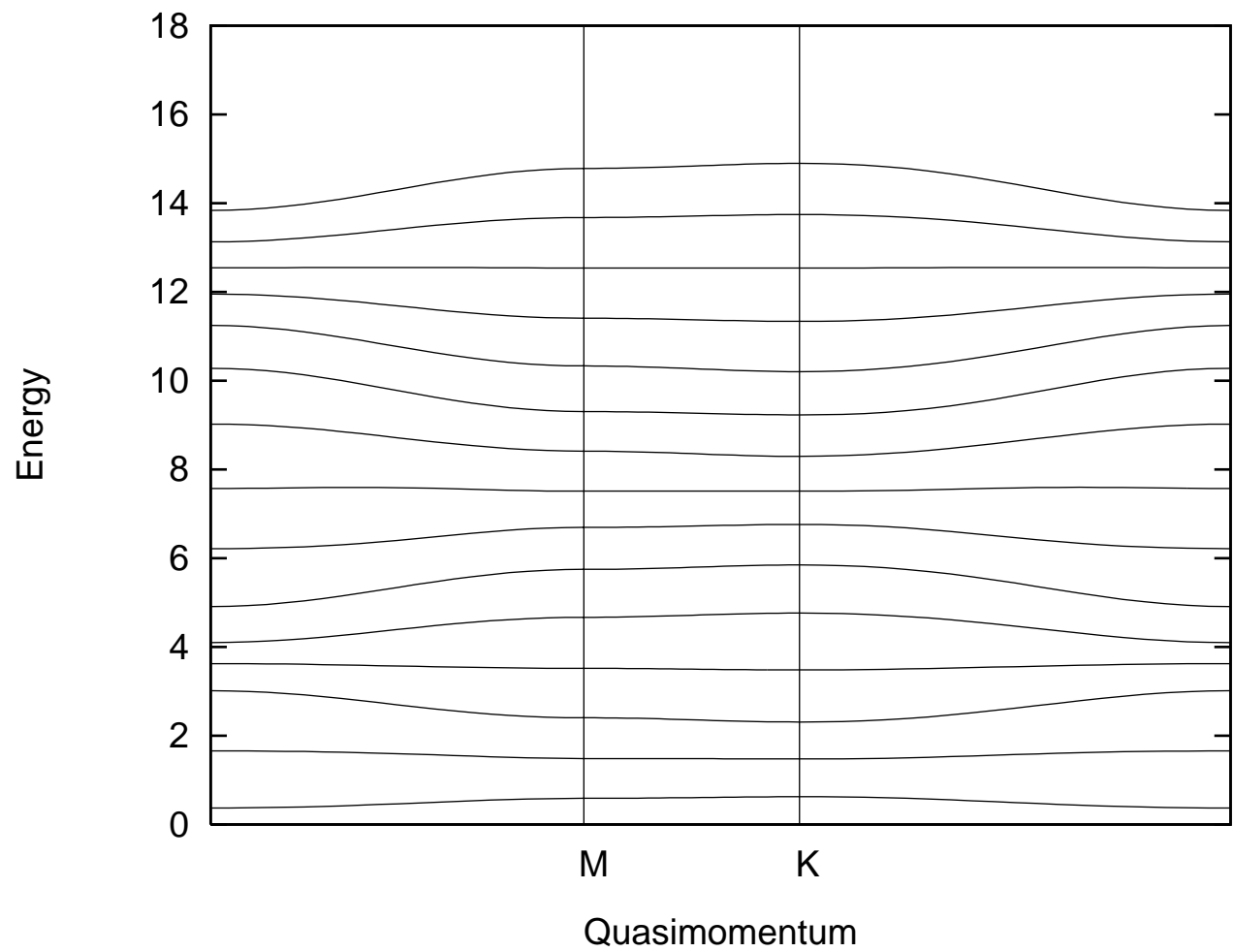
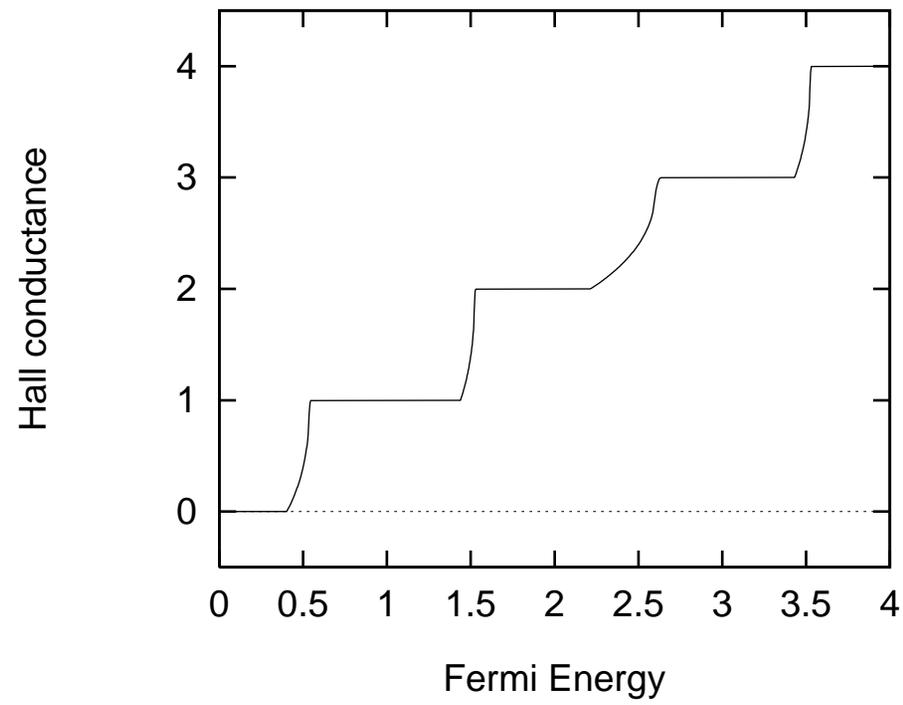
\end